\documentclass[twocolumn]{aa}
\usepackage{graphicx}
\usepackage{txfonts}
\begin{document}
   \title{Probing the atmosphere of a solar-like star by galactic microlensing at high magnification}

   \subtitle{}

   \author{F. Abe\inst{1} \and D.P. Bennett\inst{2} \and I.A. Bond\inst{3} \and J.J. Calitz\inst{4} \and A. Claret\inst{5} \and K.H. Cook\inst{6} \and 
   Y. Furuta\inst{1} \and A. Gal-Yam\inst{7} \and J-F. Glicenstein\inst{8} \and J.B. Hearnshaw\inst{9} \and P.H. Hauschildt\inst{10} \and D. Kent\inst{9} \and  
   P.M. Kilmartin\inst{9} \and Y. Kurata\inst{1} \and K. Masuda\inst{1} \and D. Maoz\inst{7} \and Y. Matsubara\inst{1} \and P.J. Meintjes\inst{4} \and M. Moniez\inst{1} \and
   Y. Muraki\inst{1} \and S. Noda\inst{1} \and E.O. Ofek\inst{7} \and K. Okajima\inst{1} \and L. Philpott\inst{12} \and N.J. Rattenbury\inst{12} \and S.H. Rhie\inst{2} \and
   T. Sako\inst{1} \and D.J. Sullivan\inst{13} \and T. Sumi\inst{14} \and D.M. Terndrup\inst{15} \and P.J. Tristram\inst{9} \and T. Yanagisawa\inst{1} \and P.C.M. Yock\inst{12}}

   \offprints{I.A. Bond, iab@roe.ac.uk}

   \institute{
   Solar Terrestrial Environment Laboratory, Nagoya University, Nagoya 464-01, Japan
   \and
   Department of Physics, Notre Dame University, Notre Dame, IN 46556, USA
   \and
   Institute for Astronomy, University of Edinburgh, Edinburgh, EH9 3HJ, UK
   \and
   Department of Physics, University of the Free State, Bloemfontein, South Africa
   \and
   Instituto de Astrof\'{i}sica de Andaluc\'{i}a, CSIC, Apartado 3004, 18080 Granada, Spain 
   \and
   Institute for Geophysics and Planetary Physics, Lawrence Livermore National Laboratory, CA 94563, USA
   \and
   School of Physics and Astronomy, Tel-Aviv University, Tel-Aviv 69978, Israel 
   \and
   CEA, DSM, DAPNIA, Centre d'\'{E}tudes de Saclay, F-91191 Gif-sur-Yvette Cedex, France
   \and
   Department of Physics and Astronomy, University of Canterbury, Private Bag 4800, Christchurch, New Zealand
   \and
   Universit\"{a}t Hamburg, Sternwarte, Gojenbergsweg 112, D-21029 Hamburg, Germany
   \and
   Laboratoire de l'Acc\'{e}l\'{e}rateur Lineaire, IN2P3 CNRS, Universit\'{e} de Paris-Sud, F-91405 Orsay Cedex, France
   \and
   Faculty of Science, University of Auckland, Private Bag 92019, Auckland, New Zealand
   \and
   School of Chemical and Physical Sciences, Victoria University, PO Box 600, Wellington, New Zealand
   \and
   Department of Astrophysical Sciences, Princeton University, Princeton NJ 08544, USA
   \and
   Department of Astronomy, Ohio State University, Columbus OH 43210, USA
   }

   \date{Received June 00, 2003; accepted July 00, 2003}

   \abstract{
   We report a measurement of limb darkening of a solar-like star in the
   very high magnification microlensing event MOA~2002--BLG--33. A 15 hour
   deviation from the light curve profile expected for a
   single lens was monitored intensively in V and I
   passbands by five
   telescopes spanning the globe. Our modelling of the light curve showed
   the lens to be a close binary system whose centre-of-mass passed almost directly in 
   front of the source star. The source star was identified as an F8--G2 main
   sequence turn-off star. The measured stellar profiles agree with current 
   stellar atmosphere theory to within $\sim$4\% in two passbands. The effective 
   angular resolution of the measurements is $<$1$\mu$as. These are the
   first limb darkening measurements obtained by microlensing 
   for a Solar-like star.
   }

   \maketitle
%
%________________________________________________________________

\section{Introduction}

In gravitational microlensing events binary lenses produce bounded regions of high
magnification, known as caustics, on the magnification map projected onto the source 
plane (Schneider \& Weiss \cite{sw}). The steep magnification gradients associated 
with these caustic curves may be utilised to resolve the surfaces of background 
source stars as they move across them. Observations of this effect have been used to
measure limb darkening in Galactic Bulge K giant stars 
(Albrow et al \cite{alb99}, Albrow et al \cite{alb00}, Fields et al \cite{fields}), 
a G/K sub-giant (Albrow et al \cite{albrow}), and an 
A dwarf in the Small Magellanic Cloud (reporting an angular resolution 
of a few nas, Afonso et al \cite{afonso00}).

In this Letter we present observations of the very high magnification 
microlens event MOA~2002--BLG--33. In this event, the centre-of-mass
of the close binary lensing system moved into near perfect alignment with a Solar-like source star.
The resulting high magnification provided ideal conditions for monitoring the source
as it transited the central caustic. Our observations yielded the most precise limb
darkening measurements obtained by microlensing for a non-giant star and the first such measurements
for a Solar-like star other than the Sun itself.

%__________________________________________________________________

\section{Light Curve of MOA~2002--BLG--33}

\begin{figure*}
\centering
\includegraphics[angle=-90,width=18cm]{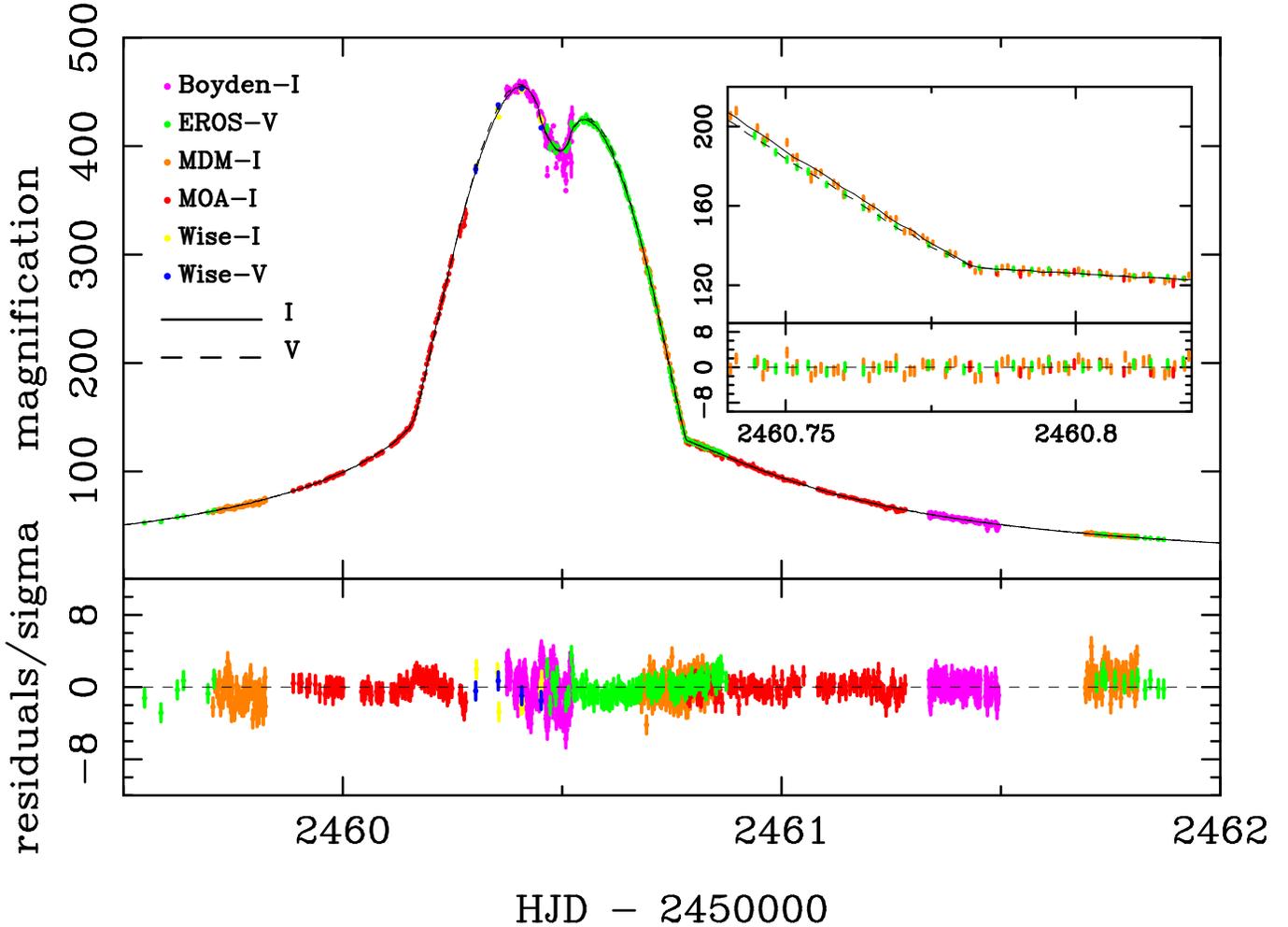}
\caption{Observed light curve, best fitting model, and residuals from the model 
for the peak of MOA~2002--BLG--33. The total dataset consists of 473 measurements in
the I band from Boyden 1.5 m telescope (South Africa), 186 points in V$_{\rm E}$ 
from EROS 1m (La Silla), 260 points in I from MDM 2.4m (Kitt Peak), 480 points in I from MOA 
0.6m (New Zealand), and 21 points in I and 10 points in V from Wise 1m (Israel).
The inset shows the separation of the light curves in the V$_{\rm E}$ and I bands 
caused by limb reddening.}
\label{fig:lc}
\end{figure*}

%% Observations
The event MOA~2002--BLG--33 was first alerted by the MOA collaboration on
2002 June 18 in its ongoing microlensing alert program (Bond et al \cite{bond01}). 
About 48 hours before its time of peak amplification on 2002 July 4, MOA
identified this event as a high magnification event in progress and a second alert was
issued. Subsequently, an intensive campaign of follow-up observations was carried 
out using a number of telescopes circling the globe (Fig.~\ref{fig:lc}). Photometry of the 
event was derived using difference imaging analysis applied to the CCD images. This procedure 
was carried out separately on
each passband of each telescope resulting in six sets of time series photometry.
These are shown in Fig.~\ref{fig:lc} where they have been normalised to
the microlensing light curve using the best fitting model (described below).

The M-shaped feature on the light curve
results from the source star entering and exiting the central caustic curve produced 
by a binary lens. The detailed shape of the light curve depends on the characteristics
of the lens, the trajectory of the source star, and the radius and limb darkening
profile of the source star. The parameters associated with the binary lens are the 
lens mass ratio $M_1/M_2$, the separation $a$ of the two lenses, the Einstein
radius crossing time $t_{\rm E}$, the source trajectory impact parameter, $u_0$, with respect
to the centre-of-mass of the lens, the time $t_0$ when the source is closest to the centre of mass of
the lens, and the angle, $\phi$, between the source trajectory and the lens components.

The source star is parametrized by its radius $r_{\rm s}$ and an
appropriate limb darkening model. For this we used the square-root law (Claret \cite{claret})
\begin{equation}
\frac{I_\lambda(\theta)}{I_\lambda(0)} = 1 - c_\lambda (1-\cos \theta) - d_\lambda (1 - \sqrt{\cos \theta})
\label{eqn:ld}
\end{equation}
where $c_\lambda$ and $d_\lambda$ are the wavelength dependent limb darkening coefficients. 

Model binary microlensing light curves were generated numerically using an inverse ray tracing 
technique, implemented on a cluster computer (Rattenbury et al. \cite{rattenbury}). The
inverse ray tracing technique naturally accommodates the
finite size of the source star and the limb darkening profiles across its 
face (Wambsganss \cite{wambs}). To find the best 
fitting model light curve, we implemented a $\chi^2$ 
minimisation technique using a Metropolis Hastings Markov chain Monte Carlo 
procedure (Ford \cite{ford}). This avoids the problem of traditional
minimisation techniques which can be fooled by false minima due to
numerical noise in the simulated light curves.

The best fitting theoretical curve is shown in Fig.~\ref{fig:lc}. A total of 11 physical
parameters plus 6 flux normalisation parameters (one for each passband-telescope set)
are required to describe the light curve\footnote{Blending parameters were not required because 
difference imaging analysis removes the effects of blending due to nearby neighbours and the
lens itself.}. The best fitting physical parameters are given in Table~\ref{tab:parameters}.
The fitting procedure yielded correlated values for the limb darkening coefficients. 
In Table~\ref{tab:parameters},
we report the coefficients ($c^\prime_\lambda$, $d^\prime_\lambda$), which are related to the
coefficients in Eqn.~\ref{eqn:ld} through the phase space rotation 
$c^\prime_\lambda=c_\lambda\cos 33\degr+d_\lambda\sin 33\degr$ and 
$d^\prime_\lambda=d_\lambda\cos 33\degr-c_\lambda\sin 33\degr$. 
In this representation, the $d^\prime_\lambda$ term contributes a small correction to
$I_\lambda(\theta)/I_\lambda(0)$.
The limb darkening parameters allow a colour map of the face of the source star to be drawn.
This is depicted in Fig.~\ref{fig:caustic} along with its passage through the caustic. 

\begin{table}
\caption{Microlensing parameters of MOA~2002--BLG--33. Here $r_{\rm E}$ denotes the Einstein radius of 
the lens, as determined by its total mass, $M_1 + M_2$, $D_{\rm OL}$ and $D_{\rm OS}$ the distances to the lens 
and source stars. The errors are 2$\sigma$ limits calculated as in Avni (\cite{avni}).
}
\label{tab:parameters}
$$
\begin{array}{p{0.3\linewidth}p{0.15\linewidth}p{0.45\linewidth}}
\hline
\noalign{\smallskip}
Parameter & & Value \\
\noalign{\smallskip}
\hline
\noalign{\smallskip}
Lens mass ratio  & $M_1/M_2$ & 0.54$\pm$0.20\\
Lens separation  & $a$ & 0.11$\pm$0.01 $r_{\rm E}$\\
Einstein time    & $t_{\rm E}$ & 50.7$\pm$1.0 day\\
Position angle   & $\phi$ & 112\degr$\pm$2\degr\\
Source radius    & $r_{\rm s}$ & (2.25$\pm$0.05)$\times 10^{-3}$ $r_{\rm E}D_{\rm OS}/D_{\rm OL}$\\
Impact parameter & $u_0$ & (4.29$\pm$0.09)$\times 10^{-4}$ $r_{\rm E}$\\
Centre time      & $t_0$ & HJD2452460.496$\pm$0.004 \\
Limb darkening   & $c^\prime_{\rm I}$ &  0.41$\pm$0.04 \\
                 & $d^\prime_{\rm I}$ & -0.08$\pm$0.65 \\
                 & $c^\prime_{\rm V_E}$ &  0.49$\pm$0.02 \\
                 & $d^\prime_{\rm V_E}$ &  1.15$\pm$0.56 \\
\noalign{\smallskip}
\hline
\end{array}
$$
\end{table}

\begin{figure}
\centering
\caption{The passage of the source star of MOA~2002--BLG--33 through the lens-caustic. 
The magnification map of the lens is represented in blue, with regions of higher 
magnification represented by lighter shading. The caustic is the diamond shaped 
line of very high magnification. The red lines indicate the track of the source star relative to the 
caustic. The position angle of the lens binary system with respect to the source trajectory is denoted by $\phi$.
The source star is shown to scale and in realistic colour. Its angular diameter would be 2--4$\mu$as.        
}
\label{fig:caustic}
\end{figure}

Our best best fitting model light curve and the associated caustic correspond
to a close binary lensing system. 
Dominik (\cite{dominik}) found that qualitatively similar caustic curves can be 
produced by lensing configurations that differ greatly. Caustics similar in
shape to the one shown in Fig.~\ref{fig:caustic} can be produced by lenses 
comprised of wide binary systems. Our best fitting wide binary solution
for  our observations of MOA~2003--BLG--33
yielded a mass ratio corresponding to that of a giant planet orbiting a stellar
mass object. While this configuration qualitatively produced a short 
duration M-shaped amplification profile, it could not reproduce the observed light curve
profile at a level demanded by the precise, densely sampled photometry we obtained.
Only the above close binary model gave an acceptable fit to the data.

Unfortunately, neither the 
rotation period of the binary lens, nor the plane of rotation, was able to be determined from the 
data. We estimated the contribution of binary rotation to our measurements of limb darkening
as follows.
For typical values of the lens parameters, the rotation 
period lies in the range 100--200 days. During the caustic crossing shown in Fig.~\ref{fig:caustic}, 
the lens is thus expected to rotate through 1\degr--2\degr.  
Light curves were computed
for rotating lenses with various periods of rotation, 
but with the plane of rotation held fixed and perpendicular to the lens-source 
axis, to maximise the effect. It was found that periods of rotation $<$150 days 
yielded light curves that were inconsistent with the data at a level greater than $2\sigma$. 
Limb darkening coefficients, and stellar profiles, were then computed, in the 
static approximation, for a simulated dataset with a rotation period of 150 days. 
It was found that the computed stellar profiles differed from the inputted profiles 
by approximately 2$\times$ the 2$\sigma$ statistical uncertainty of the profiles computed without 
rotation. The effect of rotation was therefore allowed for, in the present work, by 
doubling the errors in the stellar profiles calculated in the static approximation. 

\section{The surface of the source star}

We used colour measurements to identify the source star type in MOA~2003--BLG--33.
The baseline (i.e. unmagnified) flux of the source 
star was determined by from the fluxes measured by the MDM and Wise telescopes 
at high magnification. The instrumental magnitudes were calibrated using the catalogue 
of Udalski et al (\cite{udalski}). 
The magnitudes thus determined in the V and I bands 
(Landolt system) were $m_{\rm V} = 19.94\pm0.05$ and $m_{\rm I} = 17.96\pm0.07$. 
Assuming a distance to the source star of $7\pm2$ kpc, and absorption and reddening 
due to galactic dust with $A_{\rm I} = 1.18\pm0.05$ mag and $\mathrm{E(V-I)} = 1.24\pm0.05$ 
(Sumi et al. \cite{sumi03}, Udalski \cite{udalski03}), we derive its absolute 
magnitude $M_{\rm I}=2.6\pm0.6$ 
and colour index $\mathrm{V-I}=0.74\pm0.10$. These values identify the source 
star as an F8--G2 main-sequence turn-off star with age approximately 3--10 Gy 
and temperature in the range 6200--5800K (Girardi et al. \cite{girardi}), i.e. a solar-like 
star that has just commenced evolving away from the main sequence, or is just about 
to do so.

\begin{figure}
\centering
\includegraphics[angle=0,width=9cm]{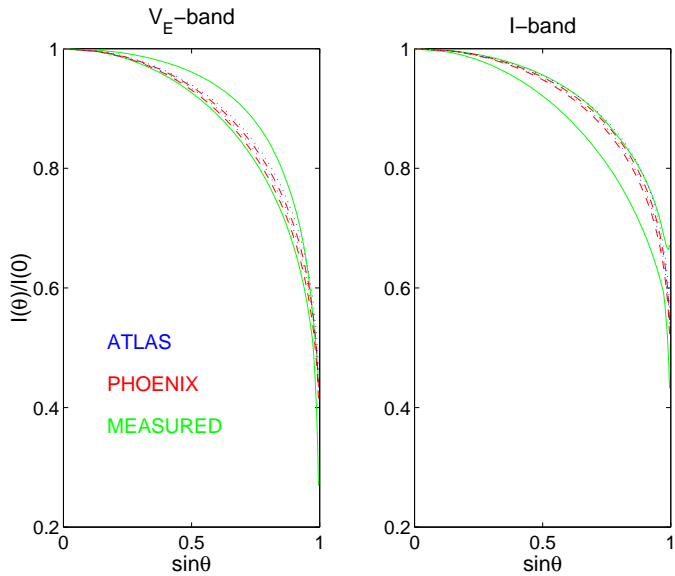}
\caption{Comparison of the measured profiles at visual (left) and infrared (right)  
wavelengths of the source star of MOA~2002--BLG--33, with predictions of the ATLAS and PHOENIX 
models of stellar atmospheres (Claret 2000). 95\% confidence limits are shown for the measured 
profiles. The x-axis variable, $\sin\theta$, is the distance from the centre of the face of the star 
normalised to the star's radius. For the model atmospheres, the temperatures have been set at 
5800K (lower curves) or 6200K (upper curves), the metallicity at $\mathrm{[M/H] = -0.3}$ 
(Tiede et al. \cite{tiede}), and the surface gravity $\log g = 4.0$.}
\label{fig:limb}
\end{figure}

Limb-darkening coefficients have been calculated for the ATLAS and PHOENIX models of 
stellar atmospheres for a wide variety of stars (Claret \cite{claret}). The limb darkening
coefficients, ($c^\prime_\lambda$, $d^\prime_\lambda$),
for the ATLAS model of a star with $T_{\rm eff}$=6000K, surface gravity
$\log g$=4.0, and metallicity $\mathrm{[M/H] = -0.3}$ are (0.363, 0.459) for the I-band
and (0.461, 0.394) for the V$_{\rm E}$-band\footnote{For the V$_{\rm E}$-band, the average
values given by Claret (\cite{claret}) in the V and R bands were used, to allow for the extended
red transmission of the EROS V filter, and also the pre-filtering effect of galactic dust.}.
These compare favourably with the measured values in in Table~\ref{tab:parameters} that were 
derived from our observations without including the effect of rotation described above.
Fig.~\ref{fig:limb} 
shows the theoretical intensity profiles for the range of possible star types identified
for MOA~2003--BLG--33, and the intensity profiles derived from the measurements with the
effect of rotation included.
It is apparent that the ATLAS and PHOENIX profiles are confirmed at the 95\% confidence level 
to within about 4\%. The precision of the present measurements resulted from the fact that the
magnification was high throughout the caustic crossing, and that it could 
therefore be monitored accurately everywhere, in particular at the end points which are most 
sensitive to limb darkening and reddening, as shown in Fig.~\ref{fig:lc}. 
The effective angular resolution of our observations of MOA~2002--BLG--33
is $<1 R_{\sun} /7$ kpc, i.e. $<$1$\mu$as.

Our results appear to be in agreement with those expected by stellar atmosphere
models for the source star type identified here. Limb darkening measurements for
another solar-like star, and consistent with those reported here, have been obtained 
by Deeg et al (\cite{deeg}) using a planetary transit technique.
The only other observations by gravitational
microlensing of limb darkening on a main sequence star were obtained for the A dwarf source
star in MACHO~1997--SMC--1 (Afonso et al \cite{afonso00}). These were only partly in
agreement with stellar atmosphere theory (Claret \cite{claret}).
Limb darkening measurements in agreement with theoretical expectations
have been reported for a G/K subgiant source in OGLE~1999--BLG--23 (Albrow et al \cite{albrow}), and for
the K giant source stars in MACHO~1997--BLG--28 (Albrow et al
\cite{alb99}) and in MACHO~1997--BLG--41 (Albrow et al \cite{alb00}). On the other hand,
recent results on source star in EROS~2000--BLG--5, also a K giant, appear
in strong disagreement with stellar atmosphere theory (Afonso et al \cite{afonso01}, 
Fields et al \cite{fields}) whereas no such discrepancy was observed in 
H$_\alpha$ equivalent width variations obtained in spectroscopic observations of the same event
(Albrow et al \cite{alb01}).

\section{Conclusion and outlook}

We have shown how a high magnification microlensing event, with the lens being a close
binary system, can be utilised to profile the atmospheres of a dwarf star. Such stars, even if nearby, 
cannot be resolved using conventional techniques. Thus the high magnification microlensing 
technique may prove useful until new, major programmes with the Very Large 
Telescope Interferometer and the Keck Interferometer become fully operational 
(Segransan et al. \cite{segransan}, Domiciano de Souza et al. \cite{domiciano}).
We note that although some previous measurements using microlensing techniques 
yielded possible evidence for significant departures from conventional stellar atmosphere 
theory, no such evidence was obtained in the 
present measurements. Indeed, as the star under study here was quite similar to the 
Sun, no such departure was expected.   

\begin{acknowledgements}
We thank the referee, J-P. Beaulieu,
for helpful suggestions in improving the presentation of this manuscript.  
We thank P. Dobcsanyi, B. Donovan, G. Gimel'farb, A. Gould, A. Gyekye, P. Leonne and S. Peale for 
assistance, the Space Telescope Science Institute for Director's Discretionary time on the HST, 
the Marsden Fund of New Zealand, the Department of Education, Science and Culture of Japan, 
the NSF (USA),  and the University of Auckland for financial support, and also the P\^{o}le 
Scientifique de Mod\'{e}lisation Num\'{e}rique (ENS-Lyon), the Norddeutscher Verbund f\"{u}r H\"{o}ch und 
H\"{o}chstleistungsrechnen (HLRN), and the NERSC for computer time. This work was performed under 
the auspices of the U.S. Department of Energy,
National Nuclear Security Administration by the University of California, 
Lawrence Livermore National Laboratory under contract No. W-7405-Eng-48.
\end{acknowledgements}

\end{document}